\documentclass[twocolumn,preprintnumbers,amsmath,aps]{revtex4-1}
\usepackage{graphicx}
\usepackage{dcolumn}
\usepackage{bm}
\usepackage{subfigure}
\usepackage{color}
\usepackage{microtype}
\begin{document}
\def\eq#1{(\ref{#1})}
\def\fig#1{figure\hspace{1mm}\ref{#1}}
\def\tab#1{table\hspace{1mm}\ref{#1}}
\title{Canonical Schottky barrier heights of the transition metal dichalcogenide monolayers in contact with a metal}

\author{Dominik Szcz{\c{e}}{\'s}niak$^{1}$}\email{d.szczesniak@ajd.czest.pl}
\author{Ross D. Hoehn$^{2, 3}$}
\author{Sabre Kais$^{2, 3}$}

\affiliation{$^1$Institute of Physics, Jan D{\l}ugosz University in Cz{\c{e}}stochowa, Ave. Armii Krajowej 13/15, 42200 Cz{\c{e}}stochowa, Poland}
\affiliation{$^2$Qatar Environment and Energy Research Institute, Hamad Bin Khalifa University, Qatar Foundation, PO Box 5825, Doha, Qatar}
\affiliation{$^3$Departments of Chemistry, Physics and Birck Nanotechnology Center, Purdue University, West Lafayette, 47907 Indiana, USA}
\date{\today} 
\begin{abstract}

The transition metal dichalcogenide ($MX_{2}$, where $M$=Mo, W and $X$=S, Se, Te) monolayers are of high interest for semiconducting applications at the nanoscale level; this interest is due to both their direct band gaps and high charge mobilities. In this regard, an in-depth understating of the related Schottky barrier heights, associated with the incorporation of $MX_{2}$ sheets into novel low-dimensional metal-semiconductor junctions, is of crucial importance. Herein, we generate and provide analysis of the Schottky barrier heights behavior to account for the metal-induced gap states concept as its explanation. In particular, the present investigations concentrate on the estimation of the charge neutrality levels directly by employing the primary theoretical model, {\it i.e.} the cell-averaged Green's function formalism combined with the complex band structure technique. The results presented herein place charge neutrality levels in the vicinity of the mid-gap; this is in agreement with previous reports and analogous to the behavior of three-dimensional semiconductors. The calculated canonical Schottky barrier heights are also found to be in agreement with other computational and experimental values in cases where the difference between electronegativities of the semiconductor and metal contact is small. Moreover, the influence of the spin-orbit effects is herein considered and supports that Schottky barrier heights have metal-induced gap state-derived character, regardless whether spin-orbit coupling interactions are considered. The results presented within this report constitute a direct and vital verification of the importance of metal-induced gap states in explaining the behavior of observed Schottky barrier heights at $MX_{2}$-metal junctions.

\end{abstract}
\maketitle
%

\section{Introduction}

Nanoscale materials offer a plethora of functionalities strongly influenced by (or even built upon) quantum effects, phenomenon not accessible at larger scales \cite{tang}. The advantages of using such low-dimensional materials become especially evident in the emerging fields of nano- and opto-electronics, where nanomaterials are employed to perform bulk-derived electronic operations both at greater efficiencies and with diminishing length scales \cite{lu, chau, wang, jariwala, akinwande, xia, mak}. Therein, a semiconducting materials constitute crucial components of the nano-scale devices due to their pivotal roles in: charge transport, logic operations, and energy conversion processes \cite{wang, jariwala, bonaccorso1, bonaccorso2, smith, koppens}. Particularly promising for the aforementioned applications are two-dimensional (2D) semiconducting crystals, which exhibit a variety of desirable properties and a relative ease of integration into the devices \cite{ferrari, qian, geim}. In this class of nanomaterials, interest has been growing in group-VIB transition metal dichalcogenide monolayers ($MX_{2}$ where $M$=Mo, W and $X$=S, Se, Te) \cite{coleman, chhowalla}, semiconducting counterparts of graphene \cite{novoselov1, novoselov2} characterized by direct band gaps and high charge mobilities \cite{ataca, rasmussen, manzeli}. Predictable applications of $MX_{2}$ monolayers in the domains of nano- and opto-electronics may lead to the design of the novel low-dimensional, flexible devices with exceptional electrostatic integrity \cite{yoon}, excellent sensing functionalities \cite{sarkar}, and adjustable optical properties \cite{mak}.

To fully benefit from the extraordinary properties of $MX_{2}$ monolayers, efficient injection of electronic charge into these nanoscale semiconductors must be ensured by quality metal contacts. Herein, two different topologies of the metal contacts to the $MX_{2}$ materials can be distinguished, namely the top and the edge geometries \cite{allain, kappera, kang}. Both of these geometries can be theoretically realized by using standard bulk electrodes or other 2D materials which exhibit metallic properties. In practice, most metal contacts to $MX_{2}$ crystals are a combination of the aforementioned scenarios \cite{allain, kang}. Apart from the type of the contact being used, nanoscale metal-$MX_{2}$ junctions suffer from the creation of tunneling barriers at their interfaces which strongly limit performance of the entire device \cite{allain}. Such barriers are predicted to have common Schottky character \cite{guo, allain} and their tuning is of crucial importance for the future design of the low-resistance and high-quality devices \cite{liu1}.

To allow for convenient control of the discussed Schottky barriers, the details of their fundamental nature must be established. In this context, recent experimental and computational results have predicted that the Schottky barrier heights (SBHs) at metal-$MX_{2}$ junctions are not greatly affected when the work function of the metal contact varies \cite{kim, monch1, guo}. Moreover, some of these studies also show that these SBHs do not follow the Schottky-Mott rule \cite{mott, schottky, monch1, guo}. Collectively, it is the first indication that the so-called metal induced gap states (MIGSs) may be responsible for the observed behavior of the SBHs in the $MX_{2}$ monolayers \cite{guo, monch1, monch2}; this is analogous to what can be found in the case of the bulk material \cite{tersoff1, tersoff2}. To unambiguously account for contributions due to MIGSs, as an explanation for the behavior of the SBHs in $MX_{2}$-based devices, it is desirable to address this issue explicitly at the level of the primary theoretical model.

Motivated by the above reasoning, this study presents a novel, direct and complementary theoretical analysis of the SBHs in metal-$MX_{2}$ junctions to develop a fundamental understanding of the underlying physics in the discussed Schottky contacts. Herein we suggest that the previously observed gap states in $MX_{2}$ monolayers \cite{szczesniak1} --- which are responsible for the tunneling of the electrons between the valence and conduction bands \cite{kane} --- can be interpreted as the MIGSs.  We investigate the energy dependent behavior of these states within a combination of complex band structure formalism with a cell-averaged Green's function technique --- as proposed by Tersoff for three-dimensional semiconductors \cite{tersoff1} --- to account for the MIGSs contributions to explain the previously discussed SBHs characteristics. This theoretical approach permits us to expand our considerations beyond the local structural restrictions, while simultaneously avoiding the limitations of the more commonly used supercell approaches. Furthermore, it is possible to analyze the influence of spin-orbit coupling (SOC) effects --- which are essential to the electronic properties of the $MX_{2}$ materials \cite{reyes, zhu} --- and prove that an MIGS-based mechanism occurs independently of the degree of approximation employed. As a result, the charge neutrality levels and the canonical Schottky barrier heights of $MX_{2}$-based devices are estimated and compared to other computational and experimental results.

\section{Theoretical Model}

The nature of MIGSs suggests that the SBHs at the metal-semiconductor junctions are directly related to the bulk electronic band edge properties of the constituent semiconductor \cite{heine, tersoff1}. Specific to $MX_{2}$ monolayers, such energy states in the vicinity of the band gap can be well described within the effective three-band model with SOC effects \cite{liu2}. Aforesaid theoretical depiction of the low-energy properties in the $MX_{2}$ materials is possible due to their specific lattice symmetry and the crystal field splitting properties \cite{liu2}. In this respect, both valence and conduction bands of the $MX_{2}$ systems are essentially modeled by the contributions coming from the $d$-type orbitals of the $M$-atoms \cite{zhu, liu3}. Herein, the band edges of the $MX_{2}$ monolayers are characterized by the aforementioned model, which is explicitly described by the following $d$-symmetry Hamiltonian including the intra-atomic SOC term \cite{liu2}:
\begin{equation}
\label{eq1}
\mathbf{H} = \mathbf{I}_{\rm 2 \times 2} \otimes \mathbf{H}_{0} +\mathbf{H}_{1}.
\end{equation}
In Eq. (\ref{eq1}) the $\mathbf{I}_{\rm 2 \times 2}$ symbol stands for the $2\times2$ identity matrix, the $\mathbf{H}_{0}$ denotes the third nearest-neighbor tight-binding Hamiltonian matrix within the $\{d_{z^2}, d_{xy}, d_{x^2-y^2} \}$ minimal basis for $M$-atoms, and $\mathbf{H}_{1}$ describes the intra-atomic SOC term, which can be written as:
\begin{equation}
\label{eq2}
\mathbf{H}_{1} = \lambda \bm{\sigma} \cdot \mathbf{L},
\end{equation}
in the spin-dependent minimal basis for the $M$-atoms, {\it i.e.}: \textls[-18]{$\{ \left| d_{z^2}, \uparrow \right>, \left| d_{xy}, \uparrow \right>, \left| d_{x^2-y^2}, \uparrow \right>, \left| d_{z^2},\downarrow \right>, \left| d_{xy}, \downarrow \right>, \left| d_{x^2-y^2}, \downarrow \right> \}$}. Moreover, in Eq. (\ref{eq2}), the scalar measure of the spin-orbit coupling strength (known as the spin-orbit coupling constant) is given by $\lambda$, whereas $\bm{\sigma}$ and $\mathbf{L}$ describe the spin and orbital angular momentum operators, respectively. Note that the employed minimal basis is imposed by the crystal symmetry of the $MX_{2}$ monolayer which allows hybridization only between the $d_{z^2}, d_{xy}$ and $d_{x^2-y^2}$ orbitals \cite{liu2}. Following this description, we here provide an analysis of the primitive unit cells containing only one $M$-type atom. Finally, the required estimations of the tight-binding and the spin-orbit coupling parameters are adopted from \cite{liu2}.

In principle, Eq. (\ref{eq1}) permits the determination of bulk electronic band edge properties for $MX_{2}$ materials in the framework of the following eigenvalue problem for the real wave vector ($\mathbf{k}$):
\begin{equation}
\label{eq3}
\left( \mathbf{H}-E\mathbf{I}_{6 \times 6} \right) \Psi = 0,
\end{equation}
which yields a complete set of the propagating electronic states at the energies ($E$) proximate to the semiconducting band gap. In this manner, the $\mathbf{I}_{\rm 6 \times 6}$ is the $6\times6$ identity matrix, and $\Psi$ denotes the total crystal wave function.

Note that Eq. (\ref{eq3}) permits the inclusion of supplementary states in $MX_{2}$ materials, described by the complex values of the wave vector $\mathbf{k}$; these complex wave vectors decay/grow as they depart from their originating propagating solution \cite{szczesniak1}. Hence, evanescent states are physically relevant only when the translational symmetry of a crystal is broken \cite{szczesniak1, reuter, jensen}. Of special interest here are the states which appear within the semiconducting band gap of the $MX_{2}$ monolayers and vary their character from donor-like (in the vicinity of the valence band) to the acceptor-like (when approaching the conduction band) \cite{szczesniak1}. In the case where $MX_{2}$ is attached to a metal, a qualitative argument that such evanescent states constitute the continuum of the metal electronic states within the semiconducting band gap can be made. That is to say, such decaying states are matched with the Bloch states of the metal when the Fermi energy of the state falls in the energy range of the semiconducting band gap; this observation is valid only in the vicinity of the metal-semiconductor interface, where the gap states can become locally metallic. Therefore, evanescent states can be interpreted as the MIGSs, analogous to predictions made by Tersoff for three-dimensional semiconductors \cite{tersoff1}.

According to the above reasoning, there should be a point in energy space where the valence and conduction bands of the $MX_{2}$ monolayer contribute equally to the MIGSs, marking the formation of a charge neutrality level (CNL). Close to the interface with a metal, such a point would correspond to the new position of the $E_{F}$ in $MX_{2}$ materials. This draws from the phenomenon of Fermi level pinning at metal-semiconductors interfaces, when space charge effects cause an energy offset (bending) of the semiconductor's band structure. In this context, the {\it canonical} or {\it universal} SBHs are calculated with respect to the CNL. As a specific example, when a semiconductor initially exhibits $n$-type ($p$-type) character the SBH is calculated as the absolute energy difference between CNL and the conduction (valence) band.

It should now be well-motivated that the determination of CNLs is pivotal in the current analysis of the SBHs in $MX_{2}$ monolayers. For this reason, it is convenient to recall that the evanescent states --- which compose MIGSs in the analyzed materials --- are responsible for the tunneling of electrons between edges of the valence and conductions bands, as shown in \cite{szczesniak1}. The tunneling amplitude facilitated by these states can be conveniently captured within the following cell-averaged Green's function \cite{tersoff1}:
\begin{equation}
\label{eq4}
G(E, \mathbf{R})= \int \left< \mathbf{r}+\mathbf{R} | G(E) | \mathbf{r} \right> d^{3}r,
\end{equation}
where vectors $\mathbf{r}$ and $\mathbf{R}$ mark the position of the origin unit cell and neighboring unit cells of interest, respectively. Note that $G(E, \mathbf{R})$ obeys the symmetry restrictions of Eq. (\ref{eq1}), {\it i.e.} $\mathbf{R} = \alpha\mathbf{a}_{x}+\beta\mathbf{a}_{y}$ for the $\mathbf{a}_{x}$ and $\mathbf{a}_{y}$ being the primitive lattice vectors of a given $MX_{2}$ monolayer; $\alpha$ and $\beta$ take on integer values. Moreover, in Eq. (\ref{eq4}), $G(E)$ denotes the forward propagating Green's function given by:
\begin{equation}
\label{eq5}
G(E)=\sum_{n} \left| n \right> \left< n \right| / \left( E + i\eta - E_{n} \right).
\end{equation}
In Eq. (\ref{eq5}), $n$ numerates the accessible eigenstates of the system, each with corresponding $E_{n}$ eigenvalues; $\eta$ is an infinitesimal positive number introduced to distinguish between the advanced and retarded Green's function. Substitution of Eq. (\ref{eq5}) into Eq. (\ref{eq4}) allows to arrive with the explicit form of the cell-averaged Green's function:
\begin{equation}
\label{eq6}
G(E, \mathbf{R})= 1/N \sum_{n \mathbf{k}} e^{i \mathbf{k} \mathbf{R}}/\left( E + i\eta - E_{n \mathbf{k}} \right),
\end{equation}
where $N$ counts each unit cell in the system and $\mathbf{k}$ is the wave-vector which varies over the entire Brillouin zone. For a given crystal direction, $G(E, \mathbf{R})$ changes its sign at energy point where the density of MIGSs is the largest. Together with the change of sign associated with $G(E, \mathbf{R})$, the corresponding energy states vary character from anti-bonding to bonding. Therefore, the described character discerned through Eq. (\ref{eq6}) suggests that the zeros of $G(E, \mathbf{R})$ mark the CNL for a given material.

According to (\ref{eq6}), proper determination of $G(E, \mathbf{R})$ is possible only with complete knowledge of all propagating and evanescent states of the system. However, due to the complex nature of the $\mathbf{k}$ vector for the decaying states, these states cannot be simply calculated from Eq. (\ref{eq3}); more advanced theoretical techniques are required to generate a complete picture of the system states. Specifically, the collection of the propagating and evanescent states --- typically referred to as the complex band structure (CBS) --- can be calculated by solving an inverse eigenvalue problem \cite{reuter, szczesniak1, chang, tsukamoto}. Herein, the CBSs of the $MX_{2}$ monolayers are calculated by the nonlinear generalized eigenvalue (NGEP) method \cite{szczesniak1}.  The NGEP method is generally well-suited for application to real-space Hamiltonians defined for unit cells with non-uniform translational vectors, as in the case of Eq. (\ref{eq1}). In the framework of the NGEP, the corresponding inverse eigenvalue problem reads:
\begin{equation}
\label{eq7}
\left(\mathbf{C}_{i}-\mathbf{C}_{i+1}\vartheta-...-\mathbf{C}_{i+j-1}\vartheta^{j-1}-\mathbf{C}_{i+j}\vartheta^{j}\right) \Psi=0,
\end{equation}
where $\mathbf{C}_{i}$ is the partial Hamiltonian matrix describing interactions within the $i^{th}$ origin unit cell. Interactions with the neighboring unit cells are given in the terms of the $\mathbf{C}_{i+j'}$ matrices; $j'\in \left< 1, 2, \dots, j-1, j \right>$), where $j=\alpha$ or $\beta$ depend on the chosen crystal direction. Within \ref{eq3}, $\vartheta=e^{{\rm i} \mathbf{k} \mathbf{R}}$ represents the generalized Bloch phase factor; $\Psi$ denotes the wave function column vector, which is a combination of the partial wave functions of all unit cells under consideration.

Equation (\ref{eq7}) cannot be solved directly, therefore a linearization method --- akin to that presented in \cite{szczesniak1} --- can be used to rewrite the initial eigenvalue problem in a linear form:
\begin{equation}
\label{eq8}
\mathbf{D}\Phi=\vartheta\mathbf{E}\Phi,
\end{equation}
where $\mathbf{E}^{-1}\mathbf{D}$ denotes the block companion matrix of the assocated nonlinear eigenvalue problem (\ref{eq7}), and $\Phi$ is the corresponding wave function column vector. For more technical details on the linearization method please see \cite{szczesniak1}.

In summary, it is vital to note that for the purpose of the present analysis the three-band model --- given by Eq. (\ref{eq1}) --- provides a trade-off between predictive capabilities and computational efficiency during numerical calculations; additional note should be made to the ease of integration with the other techniques used in this study, the possibility of local analysis at the primitive unit cell level, and the ease of direct physical interpretation of the theoretical estimates. In what follows, it allows the most fundamental insight into the underlying physics of the SBHs in $MX_{2}$ monolayers, while capturing all distinct and relevant electronic properties of these materials, {\it i.e} significant SOC on the $M$-atoms, energy band splitting at the $K$-points, the band-gap magnitude, as well as the Berry curvatures. In this context, it should also be remarked that many-body electron effects are omitted in the model Hamiltonian in accordance with observed suppression of such effects in charge transfers between the $MX_{2}$ monolayers and metal electrodes \cite{zhong}.

\section{Numerical results}

In the present study, a combination of complex band structure and cell-average Green's function techniques is employed in the framework of in-house developed numerical iterative procedures. Particularly for $MX_{2}$ monolayers, the two-variable Green's function of Eq. (\ref{eq5}) is evaluated over an energy range of the three-band electronic structure and along the $\Gamma$-M-K crystal direction in the Brillouin zone. In this manner, it is possible to conveniently search for the evanescent states within the band gaps of $MX_{2}$ crystals, which are located at the K-point \cite{szczesniak1}. Furthermore, the decaying states which penetrate the gap deepest --- {\it i.e.} the states characterized by the smallest value of the Im[k] --- are required to be found for the calculation of the CNLs. These are the states which will have decay rates large enough to constitute the continuum of the metal's Bloch states into the semiconducting band gap. To locate such states it is important to conduct calculations with large enough $\alpha$’s (note, $\alpha$ enters $\mathbf{R} = \alpha\mathbf{a}_{x}+\beta\mathbf{a}_{y}$), extending consideration beyond the local origin primitive unit cell \cite{allen}. Next, for the states of interest, the sign of the cell-average Green's function is traced to find the point(s) where $G(E, R)$=0, marking the CNL for a given $MX_{2}$ material.

\begin{figure}[ht!]
\includegraphics[width=\columnwidth]{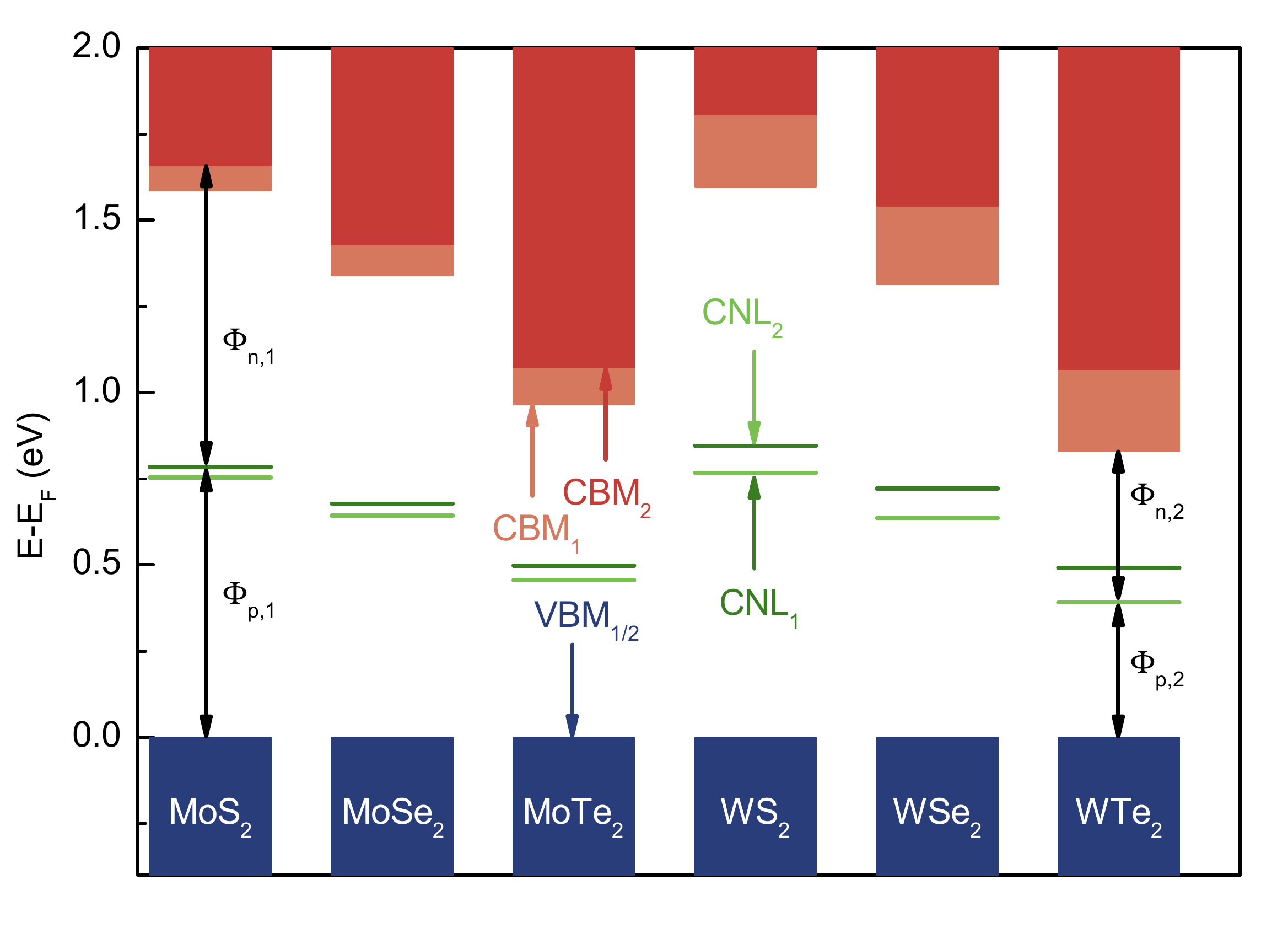}
\caption{A diagrammatic representation of the calculated CNL$_{y}$ in the analyzed $MX_{2}$ monolayers ($M$=Mo, W; $X$=S, Se, Te); where the lower index $y$ takes on values of 1 or 2, distinguishing between results obtained with and without the consideration of SOC, respectively. The energy ranges of the valence bands are marked in blue color, whereas ranges of the conduction bands are depicted by either red or light red colors for results obtained with and without SOC, respectively. For transparency, only the conduction bands minima (CBM$_{y}$) assumes different energy values, and the valence band maxima (VBM$_{y}$) are taken to be the zero-energy reference levels. The $n$- ($\Phi_{n,y}$) and $p$-type ($\Phi_{p,y}$) SBHs are noted by the black arrows.}
\label{fig01}
\end{figure}

Calculated positions of the CNLs for $MX_{2}$ monolayers are presented in the diagrammatic manner, Fig. \ref{fig01}. For convenience, the zero-energy reference level is set to the valence band maxima of the respective material; all results presented herein are given with respect to this level. Moreover, results obtained with and without the SOC effects are distinguishable, allowing one to simultaneously investigate the influence of these effects in the properties of the $MX_{2}$ crystals.

The most crucial observation that can be made on the basis of the results presented in Fig. \ref{fig01} is that the calculated CNLs lay near the mid-gap for each of the considered $MX_{2}$ materials. Such behavior suggest that the density of states (DOS) is similar both in the vicinity of the valence band maximum (VBM) and near the conduction band minimum (CBM). Indeed, as shown in \cite{yue, li} the projected DOS for the orbital make-up used here is distributed almost equally around the band gaps of the $MX_{2}$ monolayers. In this context, estimated mid-gap positions of the CNLs follow the behavior of three-dimensional semiconductors, where SHBs are induced due to MIGSs \cite{tersoff1} when electronegativities of the contact and metal are similar \cite{monch2}. Moreover, it is found that the obtained results are in agreement with previously presented density functional theory (DFT) predictions \cite{guo}, which suggest mid-gap position of the CNLs in $MX_{2}$ monolayers for top metal contact geometries. Indeed, our estimates should be interpreted as essentially valid for the aforementioned types of the interfaces. Specifically, the herein employed minimal basis --- composed of the $d_{z^2}, d_{xy},$ and $d_{x^2-y^2}$ orbitals for $M$ atoms --- favors bonding with a metal along the $z$-direction, {\it i.e.} orthogonal to the $MX_{2}$ plane.

By inspection, it is noted that the previously discussed behavior of the CNLs is obtained regardless of the inclusion of SOC effects. Therefore it can be also argued that SBHs, which correspond with calculated CNLs have, intrinsically MIGSs-derived character, despite the degree of approximation. Naturally, the positions of the CNLs differ between the two employed approaches, namely CNLs obtained by the model neglecting SOC effects (CNL$_{1}$) are slightly higher in energy than those calculated by considering SOC effects (CNL$_{2}$). In general, this effect can be attributed to a broadening of the band gap in $MX_{2}$ monolayers when SOC effects are taken into account. Additionally, the SOC effects manifest themselves in the energy difference between CNL$_{1}$ and CNL$_{2}$, which are much higher when $M=$W than $M=$Mo due to stronger spin-obit coupling for the heavier tungsten atoms \cite{zhu, liu2}.

\begin{figure}[ht!]
\includegraphics[width=\columnwidth]{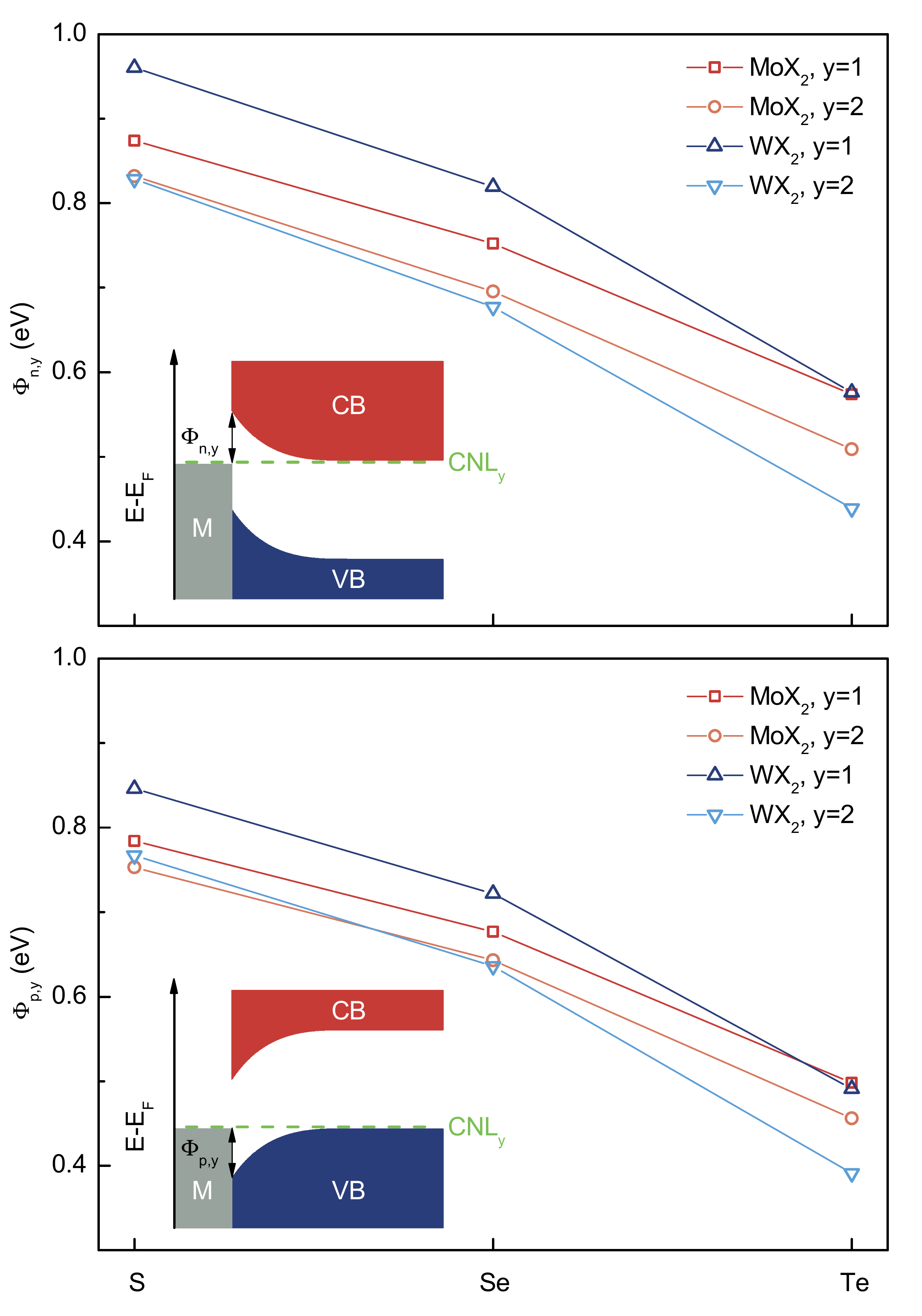}
\caption{Graphical representation of the dependence of SBHs on the constituents of the particular $MX_{2}$ monolayer ($M$=Mo, W; $X$=S, Se, Te). The $n$-type ($\Phi_{n,y}$) and $p$-type ($\Phi_{p,y}$) SBHs are depicted in the subfigure (A) and (B), respectively. For convenience, inserts qualitatively present bending of the semiconductor's bands at the metal interface, and the positions of the CNL$_{y}$ and the SBHs; where index $y\in{1,2}$ again distinguishes between results obtained with and without the inclusion of the SOC effects, respectively. Results are given with respect to the top of the valence bands, which is taken to be the zero energy reference level.}
\label{fig02}
\end{figure}

The determined positions of the CNLs permit the calculation of canonical SBHs, which arise at the metal-$MX_{2}$ material contact. According to acceptable theories of band bending mechanism for semiconductors (see schematic insets in Fig. \ref{fig02} (A) and (B)), the CNL marks point where the Fermi level of the semiconductor aligns with its metallic counterpart. Therefore, the effective potential barrier (namely the SBH) is measured as the absolute energy difference between the CNL and the CBM/VBM for the $n$-/$p$- type doped $MX_{2}$ crystal. Note, that only small degrees of electron/hole doping are assumed, such that it does not affect the electronic band edge properties in the bulk phase of the discussed materials.

The qualitative behavior of the SBHs in $MX_{2}$ materials when in contact with a metal is shown in Fig. (\ref{fig02}), subfigures (A) and (B) graphically represent the SBH-behavior for $n$-type ($\Phi_{n, y}$) and $p$-type ($\Phi_{p, y}$) doped materials, respectively. Again, the results obtained within models which neglect ($y=1$) and includes ($y=2$) SOC effects are distinguishable. The general trend --- which can be noticed in Figs. (\ref{fig02}) (A) and (B) --- is that both the $n$- and $p$-type SBHs decreases through the following chalcogen substitution S$\rightarrow$Se$\rightarrow$Te, due to a decreasing band gap size for these materials. Again note that this observation is in qualitative agreement with previously reported DFT results \cite{guo}. Moreover, it is noteworthy that the SBHs naturally follow the previously discussed behavior of the corresponding CNLs, {\it i.e.} the inclusion of the SOC effects cause decrease in the SBHs for each of the considered $MX_{2}$ materials and that SBHs are higher for most wolfram-based $MX_{2}$ materials than for those based on molybdenum. Hence, it can be provisionally stated that molybdenum-based $MX_{2}$ materials are more favorable for application in low-dimensional metal-semiconductor junctions due to lower canonical SBHs. The exact values of the calculated SBHs can be found in Table \ref{tab1}.

\begin{table}[hb!]
\caption{\label{tab1} Calculated SBHs for various $MX_{2}$ monolayers. Rows 1 and 2 give estimates of the $n$-type ($\Phi_{n}$) and $p$-type ($\Phi_{p}$) barriers without considering SOC effects. Rows 3 and 4 depict results for the $n$-type ($\Phi'_{n}$) and $p$-type ($\Phi'_{p}$) barriers while considering SOC effects. All values are given in eV.}
\begin{ruledtabular}
\begin{tabular}{c c c c c c c}
& MoS$_{2}$ & MoSe$_{2}$ & MoTe$_{2}$ & WS$_{2}$ & WSe$_{2}$ & WTe$_{2}$ \\
\hline \\
$\Phi_{n, 1}$ & 0.87 & 0.75 & 0.57 & 0.96 & 0.82 & 0.58 \\
$\Phi_{p, 1}$ & 0.78 & 0.68 & 0.50 & 0.85 & 0.72 & 0.49 \\
$\Phi_{n, 2}$ & 0.83 & 0.70 & 0.51 & 0.83 & 0.68 & 0.44 \\
$\Phi_{p, 2}$ & 0.75 & 0.64 & 0.46 & 0.77 & 0.64 & 0.39
\end{tabular}
\end{ruledtabular}
\end{table}

Based upon the results presented in Table \ref{tab1}, it was observed that the obtained values are in a good agreement with the DFT and experimental predictions of the SBHs at the interface of $MX_{2}$ crystals with the gold, ruthenium, cooper, palladium, or nickel contacts \cite{guo, monch1, kaushik}. As mentioned above, this is an expected regime of agreement between our estimates and other results, {\it i.e.} when metal contacts have top geometries and the metal-semiconductor electronegativity difference is small. In this manner, we found indications supporting the explanatory nature of MIGSs for behaviors governing the SBHs at metal-$MX_{2}$ junctions.

\section{Summary}

In the present work, SBHs of the transition metal dichalcogenide monolayers ($MX_{2}$ where $M$=Mo, W and $X$=S, Se, Te) in contact with a metal electrodes are analyzed. The investigations contained herein were conducted to account for contributions due to MIGSs to explain behaviors observed in SBHs. Contrary to previous similar studies \cite{guo, monch1}, our analysis is explicitly employed at the level of the primary theoretical model, an analogue to the investigations of the SBHs in the three-dimensional metal-to-semiconductor junctions \cite{heine, tersoff1}. In particular, we employed a combination of the complex band structure formalism and the cell-averaged Green's function technique.

The above described theoretical analysis yields estimates of the CNLs and related canonical SBHs in the devices of interest. The calculations performed suggest that the CNLs of the $MX_{2}$ monolayers are placed near the mid-point of the semiconducting band gaps. It was also observed that this behavior is in agreement with the corresponding results from density functional theory (DFT) calculations \cite{guo}, and follows the behavior of the three-dimensional metal-to-semiconductor junctions where SHBs are induced due to MIGSs \cite{tersoff1}. The obtained results are found to be particularly relevant for the top metal contact geometries and small difference in the metal-semiconductor electronegativities. Moreover, the influence of SOC effects is investigated, and prove that CNLs behave similarly regardless of whether SOC effects are included or not. In this context, we argue that the corresponding SBHs have intrinsically MIGSs-derived character despite the level of approximation used. The canonical SBHs are also found to be in agreement with both DFT \cite{guo} and experimental \cite{monch1, kaushik} predictions for the instance where the metal contacts are of: gold, ruthenium, cooper, palladium, or nickel ({\it i.e.} metals which exhibit small electronagativity differences with $MX_{2}$ monolayers \cite{monch1}). In summary, the present calculations constitute another vital test of the MIGSs contributions to explain the SBHs behavior within $MX_{2}$-metal junctions.

\section{Acknowledgement}
We would like to thank the Qatar National Research Foundation (QNRF) for their support of this project.  (Grant No. NPRP 7-317-1-055)

\bibliographystyle{apsrev}
\bibliography{bibliography}

\end{document}